\documentclass[aps,rmp,reprint,amsmath,amssymb,graphicx,longbibliography]{revtex4-1}

\usepackage{graphicx}
\usepackage[normal]{caption}
\usepackage{dcolumn}
\usepackage{bm}
\usepackage{amsmath}
\usepackage{braket}
\usepackage{eucal}
\usepackage{xcolor}
\usepackage[nodisplayskipstretch]{setspace}
\usepackage[english]{babel}
\usepackage{setspace}
\usepackage{comment}

\begin{document}
\preprint{APS/123-QED}

\title{Geometric phases of light: insights from fibre bundle theory
}

\author{C.~Cisowski}
\email{clairemarie.cisowski@glasgow.ac.uk}
\author{J.~B.~Götte}
\author{S.~Franke-Arnold}
\affiliation{University of Glasgow, School of Physics and Astronomy, Glasgow. UK.} 
\date{\today}

\begin{abstract}

Geometric phases are ubiquitous in physics; they act as memories of the transformation of a physical system. In optics, the most prominent examples are the Pancharatnam-Berry phase and the spin-redirection phase. Recent technological advances in phase and polarization structuring have led to the discovery of additional geometric phases of light. The underlying mechanism for all of these is provided by fibre bundle theory.
In this colloquium,
we review how fibre bundle theory does not only shed light on the origin of geometric phases of light, but also lays the foundations for the exploration of high dimensional state spaces, with implications for topological photonics and quantum  communications. 
\end{abstract}

%
\maketitle

\tableofcontents

\section{Introduction}

Phase is a curious protagonist in the land of physics, it bears no physical significance when a single wave is considered, yet becomes crucially important when several waves are involved, hereby causing spectacular effects such as interference. In 1984, Sir Michael Berry established that the wave function of a quantum system can gain a phase of geometric nature in addition to the dynamic phase naturally acquired over time \cite{Berry1984}. This discovery impacted various areas of physics, 
such as condensed matter, nuclear, plasma and optical physics \cite{WilczekZee1984,WilczekShapere}. Although geometric phases may appear as mere theoretical curiosities, they have led to a myriad of applications and in optics are now at the basis of wavefront shaping technologies \cite{Cohen2019,Jisha2021}. Their importance for exotic surface effects, including superconductivity and topological insulators including the quantum Hall effect, has been honoured in the recent Nobel Prize awarded to David Thouless, Duncan Haldane and Michael Kosterlitz for research on topological phases of matter.

A matrix-based formalism can be used to determine whether a system will acquire a geometric phase \cite{ONEIL200035}, however this approach gives little insight in regard to the origin of the phenomenon. Fibre bundle theory provides a deeper understanding of geometric phases: it links a phase to a state transformation based on geometrical considerations. This mathematical framework was developed 
in the first half of the 20th century, and turned out, to everyone's surprise, to provide an excellent description of gauge fields, including electromagnetic fields \cite{Yang2014}.  
It became the universal language of geometric phases almost immediately, even before Berry had time to publish his seminal work \cite{Simon1983}, and played a key role in the generalization of Berry's phase to non-adiabatic and non-cyclic systems \cite{AA1987,Bhandari1988b}.

A plethora of geometric phases have been witnessed in optics \cite{BHANDARI19971,Bhandari1988b,Vinitski1990}. Well-known examples include the Pancharatnam-Berry phase \cite{Pancharatnam1956}, born from polarization transformations, and the spin-redirection phase \cite{Rytov1938,Vladimirskiy1941}, which arises when light is taken along a non-planar trajectory. The last decades have seen significant technologocal advances in the control of phase and polarization structured light, with an ever expanding repertoire of higher order spatial modes and complex vector light fields. These developments
have revealed new geometric phases of light, caused by the transformation of spatial transverse modes \cite{VANENK199359,Galvez2003,Galvez2005,Calvo2005,Cuevas2020} and of general vectorial fields \cite{Milione2011,Milione2012,Liu:17, MosseriMilam}. However, with just a handful of exceptions \cite{Bliokh_2009,Cohen2019,Bouchiat1988} these phases are rarely linked to fibre bundles, causing key concepts such as connection and curvature to be surrounded by an aura of mathematical mystery.

In this colloquium, we illustrate how fibre bundle theory can bring about a deeper understanding of geometric phases. We do not expect the reader to have prior knowledge in the area of fibre bundle theory and will introduce a few key concepts. We show that the geometric phases recently observed in structured light beams are mostly based on two-dimensional sub-spaces of a much larger state space, and that fibre bundle theory could guide the exploration of the entire state space. Establishing a firmer link between geometric phases of light and fibre bundle theory could highlight interdisciplinary research opportunities and stimulate new discoveries. The experimental simplicity and versatility of optical systems could even allow us to test concepts of fiber bundle theory itself.   

Let us start our discussion by recalling how geometric phases differ from their dynamic counterparts using a simple interferometric construction.

\section{Geometric versus dynamic phase in a nutshell}

\begin{figure}[tb]
  \centering
  \includegraphics[width=\linewidth]{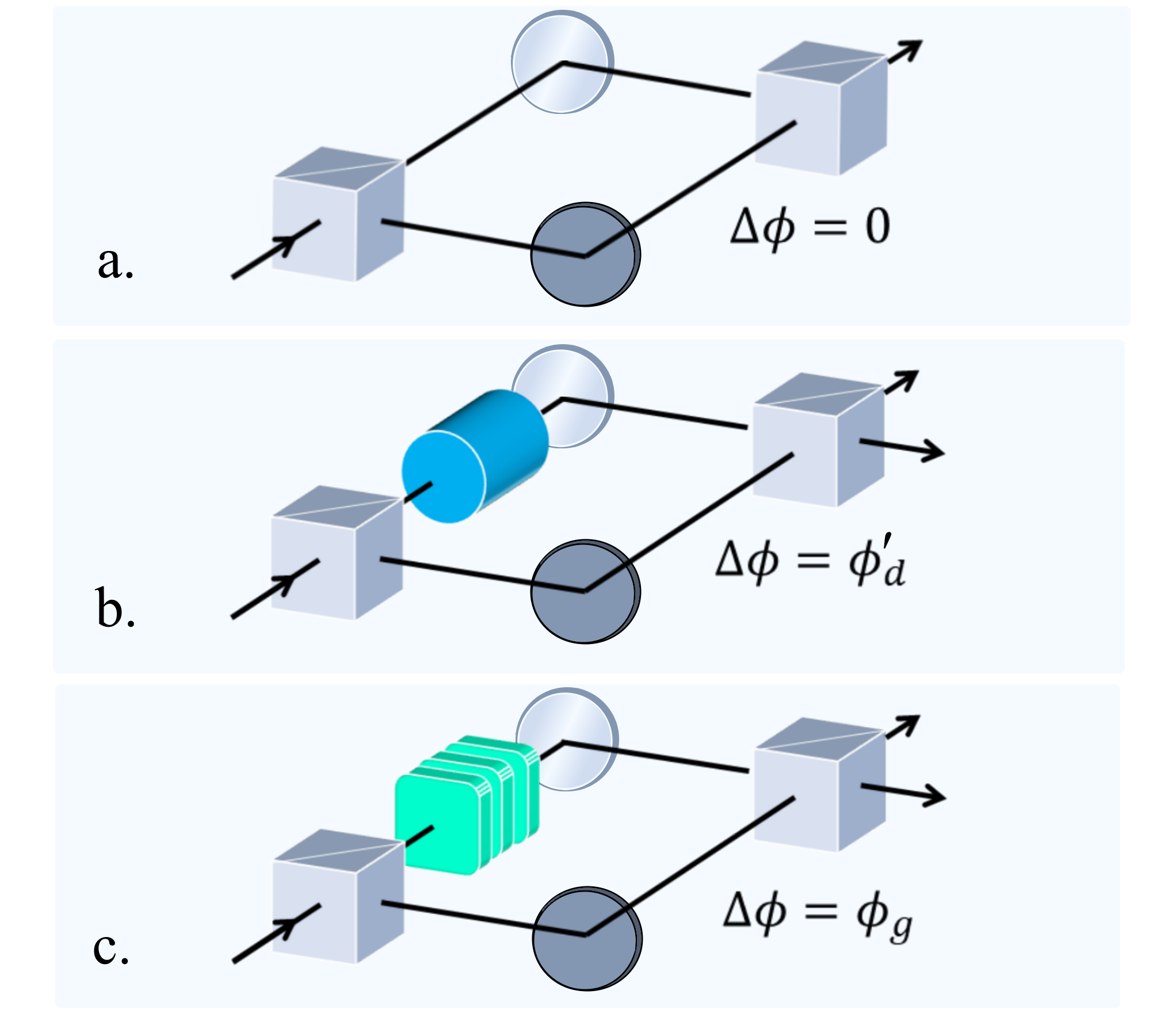}
\caption{Phase measurements with Mach-Zehnder interferometers. a. Balanced interferometer with arms of equal optical path length. b. Introducing a dynamic phase by changing the optical path length with a piece of glass in one arm. c. Introducing a geometric phase by performing a succession of state transformations in one arm.
}
\label{f1} 
\end{figure}

Interferometry is a precious tool for measuring the phase difference between two beams of light. In a Mach-Zehnder interferometer, the phase difference $\Delta\phi$ between the two beams exiting the interferometer is null if the arms of the interferometer are of equal optical path length (see Fig.\ref{f1}.a).
The phase naturally acquired over time as the beam propagates is called the dynamical phase, $\phi_{d}$. Increasing the optical path length of one of the arms, by introducing a piece of glass for instance,
will create an excess of dynamical phase $\phi_{d}+\phi'_{d}$ in this arm such that $\Delta\phi=\phi'_{d}$ (see Fig.\ref{f1}.b), modifying interference and leading to a difference in the interferometer output. 

It is also possible to obtain a finite phase difference even if the arms are of equal optical path length, by imposing a series of state transformation to one of the beams (see Fig.\ref{f1}.c) \cite{AA1987}. These transformations will cause the beam propagating through this arm to acquire a phase, $\phi_{g}$, solely dependent on the path formed in the state space, in addition to the dynamical phase acquired upon propagation, such that, at the exit of the interferometer, $\Delta\phi=\phi_{g}$. The phase $\phi_{g}$ is said to be geometric. In the next section, we show how a succession of polarization transformations can create such a geometric phase called Pancharatnam-Berry phase. 

\section{The Pancharatnam-Berry phase}

The Pancharatnam-Berry (PB) phase is one of the most ubiquous geometric phases of light \cite{BHANDARI19971,DeZela2012,Vinitski1990,LeeTan2017}. It was discovered by Pancharatnam in 1956, upon generalizing the notion of interference for partially orthogonal polarized beams \cite{Pancharatnam1956} and was identified as a geometric phase by Ramaseshan and Nityananda in 1986 \cite{Ramaseshan1986}. This led Berry to provide a quantum interpretation of this phenomenon, causing his name to be linked to this phase along with Pancharatnam's \cite{Berry1987}. 

\subsection{Experimental realisation}

The PB phase is generated by changing the polarization state of a beam of light propagating along a fixed direction. In practice, a sequence of polarization transformations can be realized using several retarders, which would correspond to the optical elements in Fig.\ref{f1}.c. For simplicity, we assume that the retarders do not change the optical path length. If the beam of light is initially horizontally polarized (state 1), we can use a quarter waveplate (QWP) to convert the beam into a circularly polarized state (state 2), use a second QWP to return the polarization state to linear (state 3), however rotated by $45^{\circ}$ with respect to the horizontal, then employ a suitably oriented half waveplate (HWP) to restore the polarization direction to horizontal (state 4). The sequence of polarization transformation is illustrated in Fig.~\ref{f2}.a. As stated previously, a geometric phase is dependent on the path formed in the state space. In order to determine whether our sequence of state transformation will generate a geometric phase, we therefore need to turn to geometric considerations. 

\subsection{Geometric interpretation}

\begin{figure}[b]
  \centering
  \includegraphics[width=\linewidth]{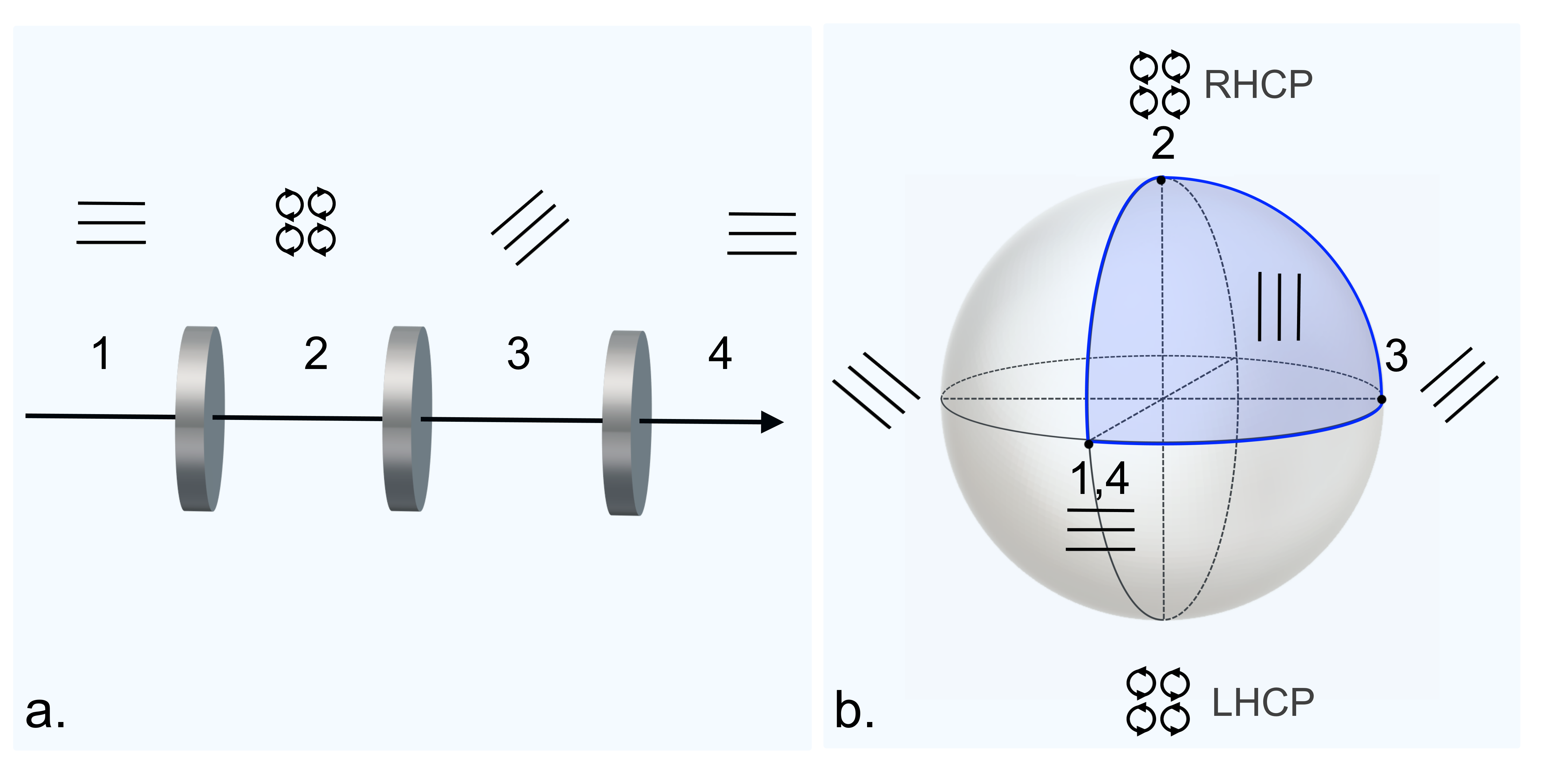}
\caption{A sequence of polarization transformations. a. Practical realisation. b. Geometric interpretation: a closed path is traced on the Poincar\'e sphere.}
\label{f2}
\end{figure}

Realizing a sequence of (unitary) polarization transformation can be visualized as a path on the Poincar\'e sphere. The Poincar\'e sphere is the state space of purely polarized light, meaning that each point on the sphere represents a pure polarization state. By convention, the poles represent circularly polarized light, the equator linearly polarized light and the hemispheres right and left elliptically polarized light (see Fig.~\ref{f2}.b). All states on the sphere can be conveniently obtained from a linear superposition of diametrically opposed states. 

The path corresponding to the polarization transformation shown in Fig.~\ref{f2}.a is drawn in Fig.~\ref{f2}.b., where successive polarization states have been linked using geodesics.

In optics, it is common practice to calculate the PB phase, which we denote $\phi_{g}$, directly from the solid angle $\Omega_{\rm PS}$ enclosed by the path formed on the Poincar\'e sphere, shown in light blue (shaded area) in Fig.~\ref{f2} \cite{Pancharatnam1956,Bhandari1988b}, using the simple relation:
\begin{equation}\label{PB}
\phi_{g}=-\frac{1}{2}\Omega_{\rm PS}.    
\end{equation}
If the sequence of polarization transformations is associated with a vanishing solid angle, no PB phase will be generated. Eq.~\ref{PB} provides a straightforward manner to calculate the PB phase, however, the relationship between phase and a path formed on the state space is far from obvious. Indeed,  in physics, states are defined up to a phase factor, meaning that two state vectors $\ket{\psi}$ and $\mathrm{exp}(i\phi)\ket{\psi}$, where $\phi\in[0,2\pi[$, are considered to be physically equivalent. A path traced on the Poincar\'e sphere thus does not directly provide information on the evolution of the phase of the system. An additional structure, capable of tracking this evolution is needed, and this is where fibre bundle theory comes into play. 

\section{The origin of geometric phases}

In what follows, we introduce some fundamentals of fibre bundle theory and show how geometric phases are interpreted in terms of fibre bundles. We then examine the PB phase from this new perspective.  

\subsection{Fibre bundle theory, a universal model}

In the landscape of mathematics, fibre bundle theory sits at the crossroads of differential geometry, topology and connection theory.
It was developed independently from physics in the first half of the 20th century \cite{Seifert1933,Hopf1931,Feldbau,Whitney1935,steenrod1951,Ehresmann,Serre1951}. The overlap with physics became evident only in retrospect, when examining 
Dirac's theory on magnetic monopoles \cite{Dirac} from a geometric perspective showed that Dirac had described a fibre bundle \cite{Lubkin1963,WuYang1975}.
Wu and Yang went further by demonstrating that fibre bundle theory is the natural language of gauge theory. They summarized this idea in a table showing how both theories describe the same concept using different terminologies 
\cite{WuYang1975}. At this point, fibre bundle theory ceased to be an abstract framework, but became suitable for the description of physical reality. This discovery profoundly influenced the physics and the mathematics community during the late 20th century, as detailed in the overview provided in \cite{Boi2004}.

\begin{figure}[b]
  \centering
  \includegraphics[width=\linewidth]{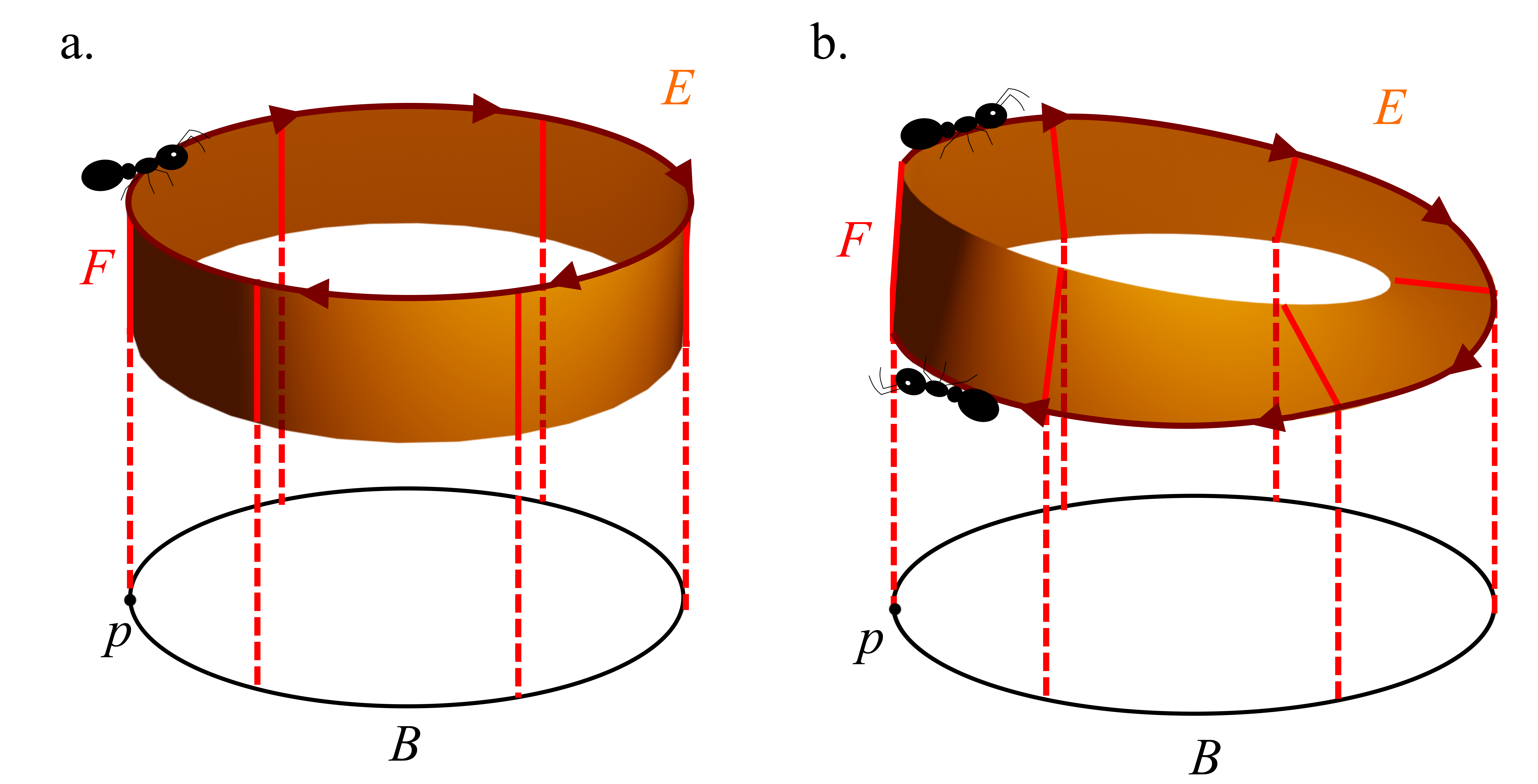}
\caption{Illustration of a trivial (a) and a non-trivial (b) fibre bundle. The base space $B$ are identical circles, several fibres are indicated by solid line segments, embedded into the total space $E$ (cylinder in a. and M\"obius strip in b.).}
\label{fnew}
\end{figure}

We illustrate the basic idea of fibre bundles in Fig.~\ref{fnew} on the example of a cylinder and a M\"obius strip.
A fibre bundle is constructed from a (topological) space $B$ called the base space (which for both Fig.~\ref{fnew}a, and b is a circle). Above each point $p \in B$ is a space called the fibre $F$ (shown as a line segment, defined by its two endpoints), linked to the base space by a projection map (indicated by dashed lines).
At the core of fibre bundle theory is the idea that locally, within a close neighborhood of points $p \in B$, the total space of the fibre bundle is the direct product of the fibre space and the base space: $E\approx B\otimes F$. For the cylinder this is true globally, making it a trivial fibre bundle. The topological and geometrical properties of the fibre bundle may however prevent us from a consistent global mapping. This is the case for the twisted M\"obius strip, as indicated by the ant, which can move from one end of the line segment $F$ to the other when traversing along the M\"obius band.  
The fiber bundle is then said to be non-trivial \cite{Batterman2002}. A fibre bundle thus contains two types of information: local and global. Physical phenomena are often studied from a local (infinitesimal) perspective; fibre bundle theory invites us to take a step back and look for global properties that may also affect our system.

To complete the definition of a fibre bundle, one may also specify the structure group $G$ acting on the bundle. In the case of the M\"obius strip, $G=\pm 1$ where the element $-1$ acts on the fibre by sending an element from the top to the bottom of the fibre \cite{Batterman2002}.

Fibre bundles relevant in optics and quantum optics tend to operate in larger state spaces, e.g. with fibres keeping track of phase evolutions. While it is perfectly feasible to describe optics in terms of complex electromagnetic fields, and quantum optics in terms of quantum states, fibre bundle theory offers a supplementary geometric interpretation, transcendending specific applications and potentially allowing us to develop a more intiuitive understanding of the underlying phenomena. 

\begin{figure}[h!]
  \centering
  \includegraphics[width=0.7\linewidth]{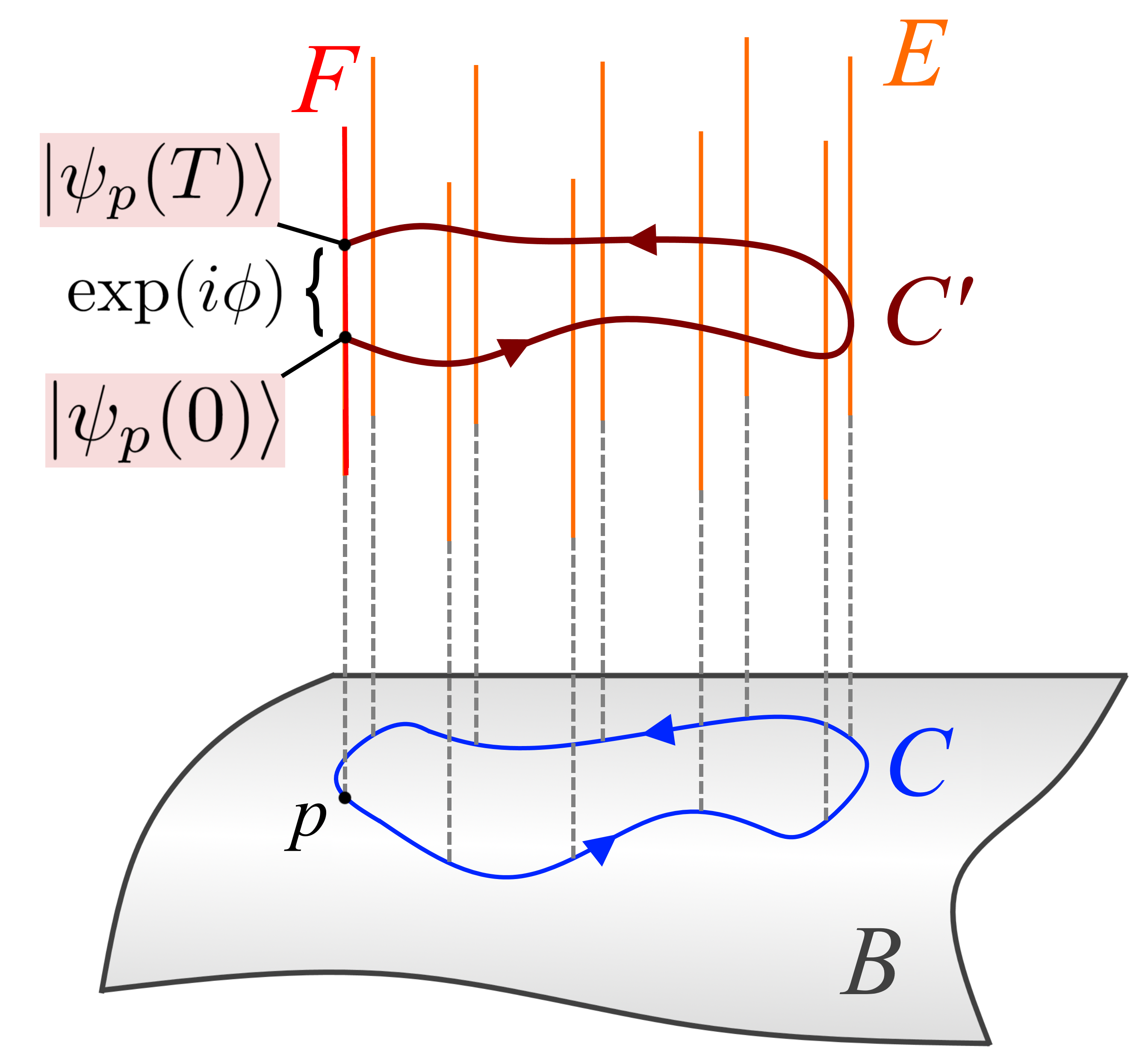}
\caption{Illustration of Aharonov and Anadan's fibre bundle. A closed path $C$, starting and ending at point $p \in B$, is lifted into the total space $E$.
The beginning and end points of $C^{'}$ lie on the same fibre $F_p$, and are related by a phase factor in Aharonov and Anadan's fibre bundle.}
\label{f3}
\end{figure}

The concept of a fibre bundle linking a phase to a state transformation was introduced by Aharonov and Anandan \cite{AA1987}. 
Here, the base space $B$ of the fibre bundle is the complex projective Hilbert space, as illustrated Fig.~\ref{f3},  which we call state space. The fibre above each state 
consists of all the normalized state vectors capable of representing that state, namely, $\exp(i\phi)\ket{\psi}$, where
$\phi\in [0,2\pi [$. The structure group is the unitary group $\text{U}(1)$ and the total space, $\text{E}$, is the Hilbert space.
This fibre bundle is a principal bundle, meaning that the fibre is the structure group.

As we mentioned earlier, a state transformation can be visualized as a path in the state space. If we assume that the transformation is cyclic, meaning that the state transforms back to the initial state at the end of the transformation, it forms a closed path $C$ in the state space, shown by the curve in the base space $B$ in Fig.~\ref{f3}. 
Knowledge of this path alone does not include any phase information.
To record phase information, the path $C$ is ``lifted" to form a path $C^{'}$ in E.
Since many curves $C^{'}$ can project down onto the same curve $C$ there are many ways to realize this lift. Some rules must be provided, this is the role of the connection $\mathcal{A}$. 
The connection decomposes the tangent space of the bundle into a vertical and horizontal component, specifying the direction along the fibres and ‘perpendicular’ to the fibres, respectively. The curve $C$ is then lifted along the horizontal direction (along the fibre in Fig.\ref{f3}). This allows us to compare (connect) points on different fibres. The splitting of the space into its horizontal and vertical components hence the connection is tied to a particular set of coordinates.

Lifting a closed path in $B$ will often result in an open path in $E$. If we assume that the beginning and end point of the path lie on the same fibre, they are linked 
by a simple phase factor $\mathrm{exp}(i\phi)$ (a U(1) transformation), called the ``holonomy of the connection on the fibre bundle" \cite{Nakahara:2003}. This phase factor indicates that the wave function has failed to come back to itself at the end of the transformation. Explicitly, for a cyclic evolution of period $T$,  $\psi_{\mathrm{p}}(T)=\mathrm{exp}(i\phi)\psi_{\mathrm{p}}(0)$. 

In the classical realm, holonomies can take the form of a rotation, such as the one allowing a cat falling from an upside-down position to land on its paws \cite{Montgomery1993} or the rotation of the oscillation plane of Foucault’s pendulum, after a day has elapsed \cite{Hannay1985,Bergmann2007}.

Aharonov and Anandan identified the connection $\mathcal{A}^\text{AA}$ that yields the geometric phase $\mathrm{exp}(i\phi)=\mathrm{exp}(i\phi_{g})$ as its holonomy. To do so, they defined the geometric phase as the difference between the total phase and the dynamic phase \cite{AA1987, Zwanziger1990}.
Of course not all evolutions are cyclic, Samuel and Bhandari showed that the path formed in the projective space can simply be closed using the shortest geodesic \cite{Bhandari1988b}, a process which does not affect the geometric phase \cite{Benedict1989}. The geometric phase can then be calculated using the connection of Aharonov and Anandan:
\begin{equation}\label{eqc}
\phi_{g}=\oint_{C} \mathcal{A}^\text{AA}=i\oint_{C}\bra{\tilde{\psi}}\mathrm{d}\ket{\tilde{\psi}},
\end{equation}
where $\mathrm{d}$ is an exterior differential operator and $\ket{\tilde{\psi}}$ is a basis vector field, also known as section or gauge (an explicit derivation can be found in \cite{Bohm2003}). Eq.~\ref{eqc} can be used if $\mathcal{A}^\text{AA}$ is uniquely defined over the region of the state space covered by the path $C$. In practice, several $\mathcal{A}^\text{AA}$ may coexist due to the geometry of the Hilbert space and of the projective Hilbert space \cite{Urbantke1991}. One may then prefer the following expression, obtained using Stokes theorem:
\begin{equation}
\phi_{g}=\int_{S} \mathrm{d}\,\mathcal{A}^\text{AA}=\int_{S} \mathcal{V}^\text{AA},
\end{equation} 
where $S$ is the surface in the state space enclosed by $C$, and $\mathcal{V}^\text{AA}$ is the curvature of the connection. Unlike the connection $\mathcal{A}^\text{AA}$, the curvature $\mathcal{V}^\text{AA}$ is well defined everywhere. It measures the dependence of the phase holonomy on the path formed in the projective Hilbert space. Geometric phases owe their name to this path dependence, and one may say that it is the curvature of the state space 
that gives birth to geometric phases \cite{Anandan1988b}. 

It is possible to witness a phase holonomy even if the curvature vanishes, when the path $C$ cannot be shrunk to a point. This typically happens if the path encloses a topological defect. The phase holonomy becomes a signature of the defect and is insensible to the shape of the path, and hence is called a ``topological phase" \cite{LYRE201445}. 

A remarkable fact about the fibre bundle interpretation of geometric phases proposed by Aharonov and Anandan is that it remains valid regardless of the dimension of the state space. We will come back to this after providing a fibre bundle interpretation of the PB phase. 

\subsection{From Poincar\'e to Hopf}

\begin{figure}[tb]
  \centering
  \includegraphics[width=\linewidth]{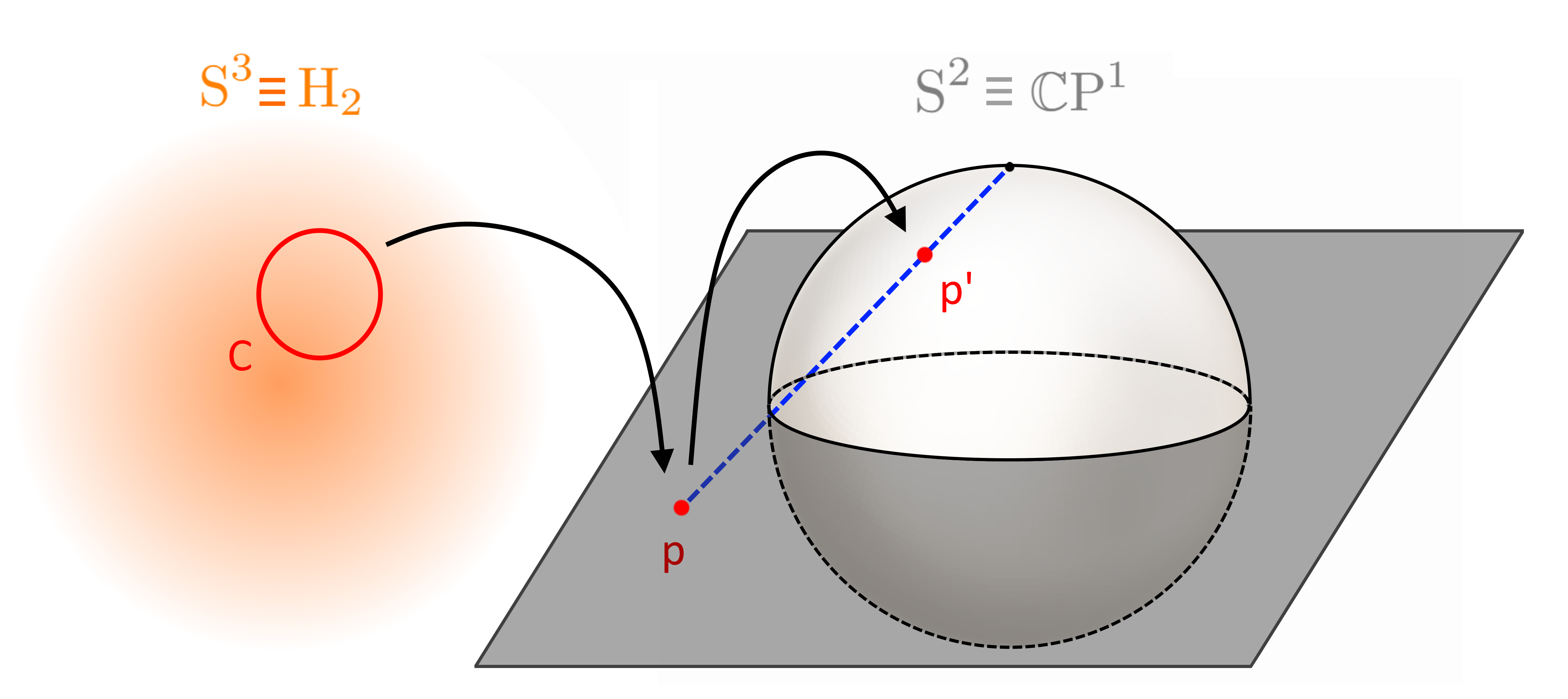} 
\caption{Schematic illustration of the Hopf map. A set of equivalent state vectors representing a pure quantum state (circle in the Hilbert space $\mathrm{H}_{2}$) is mapped onto a point in $\mathbb{R}^{2}(+\infty)$, and from there onto a point in the complex projective Hilbert space $\mathbb{C}\text{P}^{1}$ via an inverse stereographic projection (dotted line). We have chosen the North Pole as the projection point.} 
\label{F7} 
\end{figure}

In the case of the PB phase, the relevant state space is the Poincar\'e sphere. The sphere representation is specific to two-dimensional systems, and studying its construction reveals the associated fibre bundle. When fully polarized light propagates along a fixed direction, say $z$, it becomes analogous to a two-state (qubit) system:
\begin{equation}\label{eq0}
 \ket{\psi}=\alpha\ket{0}+\beta\ket{1},   
\end{equation}
where $\ket{0}$ and $\ket{1}$ are the eigenstates of the Pauli spin operator $\sigma_{z}$, and $\alpha$ and $\beta$ are complex parameters with $\mid\!\alpha\!\mid^{2}+\mid\!\beta\!\mid^{2}=1$ to ensure normalization. The state vector $\ket{\psi}$ lives in the two-dimensional Hilbert space, denoted by $\text{H}_{2}$. This space is our total space $E$, which can be pictured as a hypersphere $\text{S}^{3}$  embedded in $\mathbb{R}^{4}$, represented in orange (left shaded area) in Fig.\ref{F7}.a.
In \cite{Sugic2021} this space has been coined the \emph{optical hypersphere}. 

In optics, a pure state $\ket{\psi}$ can only be identified up to a phase factor $\exp (i\phi)$, and in quantum theory the set of state vectors $\exp (i\phi)\ket{\psi}$ describe the same physical state: the probability of a measurement of the system given by Born's rule for $\ket{\psi}$ and $\exp (i\phi)\ket{\psi}$ is one.
To account for this, in the projective Hilbert space all states $\exp (i\phi)\ket{\psi}$, where $\phi\in[0,2\pi[$, represent the same quantum state. 

This set of equivalent state vectors form a fibre, which can be pictured as a circle $\text{S}^{1}$ parametrized by $\phi$ (C in Fig.~\ref{F7}). 
For a two-state system, the state space is the projective Hilbert space  $\mathbb{C}\text{P}^{1}$, which is an ordinary sphere, known as $\text{S}^{2}$ by mathematicians. The state space is obtained by mapping each quantum state (circle) in the total space onto a point on the sphere. This mapping is performed by the Hopf map, which maps a circle onto a point $\textbf{p}$ in a plane $\mathbb{R}^{2}(+\infty)$, then maps this point onto a point $\textbf{p'}$ on the sphere via an inverse stereographic projection, as illustrated in Fig.~\ref{F7}.a \cite{Mosseri_2001}. This is how the Poincar\'e sphere, and all spheres representing two-state systems, are constructed. 

\begin{figure}[tb]
  \centering
  \includegraphics[width=\linewidth]{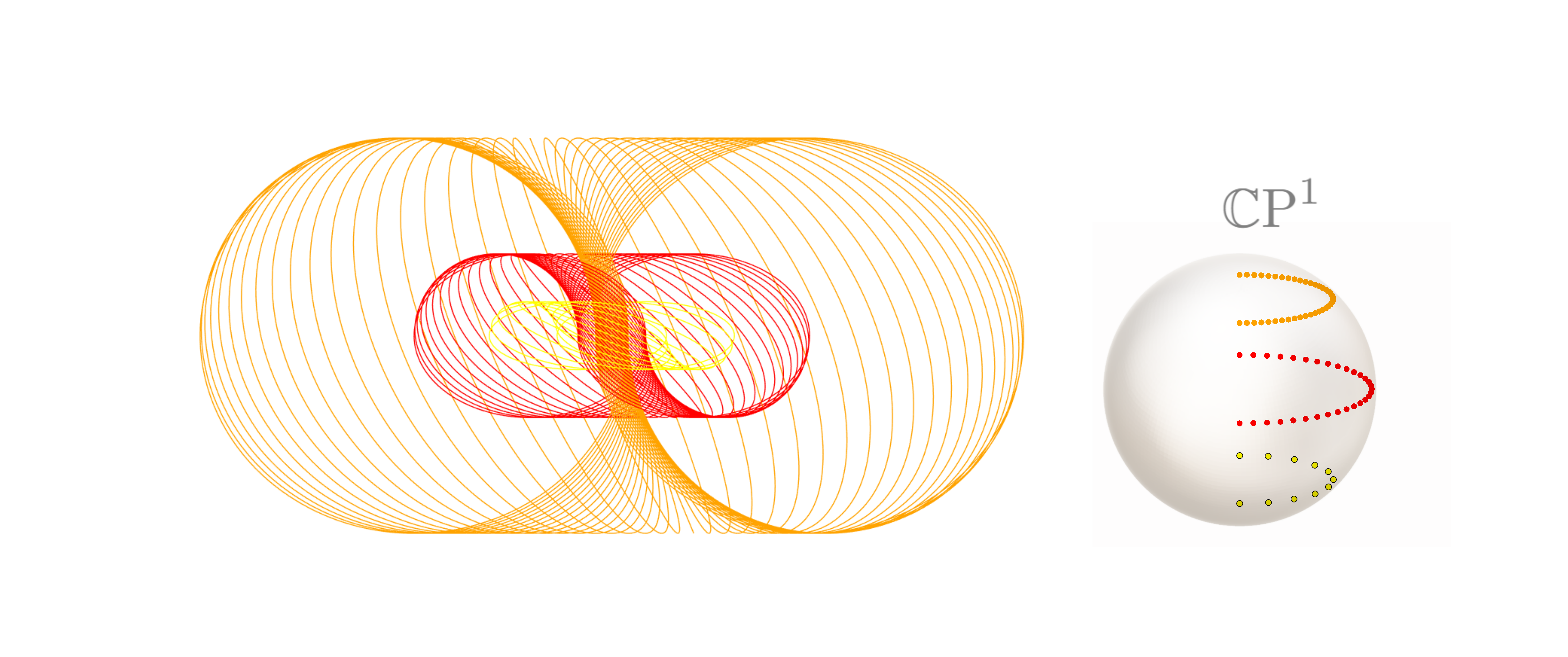} 
\caption{ Fibre structure of the Hopf fibration in $\mathbb{R}^3$. Each fibre is a circle. The set of fibres linked to the red (dark gray), orange (gray) and yellow (light gray) points on the $\mathbb{C}\text{P}^{1}$ sphere form three tori.} 
\label{F7b} 
\end{figure}

The PB phase then corresponds to the holonomy of the connection $\mathcal{A}^\text{AA}$ on a fibre bundle where the base space is $\mathbb{C}\mathrm{P}^{1}$ (the Poincar\'e sphere), a fibre is a set of equivalent states vectors, the group structure is U(1), and the total space is $\mathrm{H}_{2}$. This fibre bundle is known as the Hopf fibration, and it is capable of describing all two-state systems, not just polarization. As such, it is often encountered in physics, where it describes magnetic monopoles, two-dimensional harmonic oscillators, Taub-NUT space (relevant in the framework of general relativity) and twistors \cite{Urbantke1991}. Because it involves spaces embedded in different dimensions, the Hopf bundle is difficult to visualize, however, performing a direct stereographic map from $\mathrm{S}^{3}$ to $\mathbb{R}^{3}$ will make its fibre structure apparent \cite{mosseri_ribeiro_2007}. A schematic illustration is provided in Fig.~\ref{F7b}. 

It is not possible to assign a single connection $\mathcal{A}^\text{AA}$ over the whole Poincar\'e sphere. Indeed, if we introduce polar coordinates $\theta,\phi$, we may define a connection $\mathcal{A}$ using the basis $\ket{\tilde{\psi}}$ \cite{Kataevskaya1995,Bouchiat1988}:
\begin{equation}
\ket{\tilde{\psi}}=\left(\cos(\theta/2),e^{-i\phi}\sin(\theta/2)\right);   
\quad
\mathcal{A}=\frac{1}{2}(1-\cos\theta)d\phi  
\end{equation}
$\mathcal{A}$ is defined everywhere except at $\theta=0$ (north pole). To cover the entire sphere we may introduce another basis $\ket{\tilde{\psi'}}$ and a second connection $\mathcal{A'}$: 
\begin{equation}
\ket{\tilde{\psi}^{'}}=\left(e^{i\phi}\cos(\theta/2),\mathrm{sin}(\theta/2)\right); 
\quad
\mathcal{A'}=\frac{1}{2}(-1-\cos\theta)d\phi
\end{equation}
$\mathcal{A'}$ is defined everywhere but at $\theta=\pi$ (south pole). The singular points $\theta=\pi$ and $\theta=0$ correspond to Dirac string singularities \cite{Dirac,Yang1996}, they can be moved around the sphere by choosing different basis but cannot be removed. The sphere is thus divided into two overlapping regions, each region having a different connection. In the overlapping region, the connections are related by a phase transformation \cite{Urbantke1991}. In this case, it is preferable to calculate the PB phase using Stokes theorem:
\begin{equation}\label{sal}
\phi_{g}=\frac{1}{2}\oint_{C}(\pm 1-\cos\theta)d\phi=\frac{1}{2}\int_{S}\mathrm{sin}\theta d\theta d\phi=\frac{1}{2}\Omega_{S},  
\end{equation}
where $\Omega_{S}$ is the solid angle enclosed by the path $C$ formed in clockwise fashion on the Poincar\'e sphere. We have recovered Eq.~\ref{PB} using fibre bundle theory, and have detailed how the PB phase arises from a state transformation. Eq.~\ref{PB} is only valid because the state space can be represented as a sphere, which is true for all two-state systems. Eq.~\ref{sal} shows that the curvature on the Hopf fibration does not vanish, it confirms that the PB phase depends on the path traced in the state space, it is truly a ``geometric" phase. 

PB phases are at the core of state-of-the-art wavefront shaping technologies \cite{Bomzon02,Kim:15,Dorrah,Radwell2016}. 
Q-plates \cite{Marrucci2006}, in particular, rely on PB phases to impart a helical phase profile to a beam of light and have been commercialized under the name of vortex retarders. The underlying principle of these devices is that a spatially variant PB phase profile can be obtained by realizing spatially resolved polarization transformations.
Conversely, exotic polarization distributions such as the ones witnessed when interfering multiple beams of light \cite{Galvezpoincare,Cardano13} or obtained upon tight focusing \cite{Bauer2015} may contain interesting geometric phases profiles.

In this section we have directed our attention to fully polarized light for didactic purposes. Note however, that geometric phases can also arise from the transformation of partially polarized light. In this case, the state space becomes the Poincar\'e ball to include points inside the sphere \cite{Sjovist2000,Ericsson2003}. The geometric phase can then be obtained by purifying the state \cite{Milman2006}. Interestingly, the Poincar\'e ball naturally incorporates some hyperbolic geometry \cite{Ungar2002}, whose relevance with regard to special relativity has been highlighted in \cite{L_vay_2004} and \cite{Samuel1997}.

\section{Exploring high dimensional state spaces}

The state space of an $n$-state system where $n>2$ can no longer be represented by an ordinary sphere \cite{bengtsson_zyczkowski_2006}. Such spaces have recently become accessible in optics, through spatial transverse modes, strongly focused light and general vectorial light.
In the following we review how geometric phases are currently calculated on these spaces, and discuss how fibre bundle theory could lay the foundation for the exploration of high dimensional state spaces. Our first encounter with a high dimensional state space stems from the study of polarized beams of light with spatially varying propagation direction. 

\subsection{The spin-redirection phase}

The spin-redirection phase is a geometric phase that arises when polarized light is taken along a non-planar trajectory. 
It was first witnessed in inhomogeneous media \cite{Rytov1938,Vladimirskiy1941} and in optical fibres \cite{Ross1984, ChiaoWu1986}, in which case it is the result of an adiabatic transformation, meaning that a photon that is initially in an eigenstate of the spin operator, aligned with the direction of the wave vector, will remain in this eigenstate at all times.  In other words, its helicity does not change upon propagation. 
At the time, it was believed that the cycling of the parameters driving the adiabatic transformation determines the existence of geometric phases. Geometric phases were calculated from the path traversed in time formed in the space of parameters, in our case, the sphere of directions of the wave vector $\bm{R}(t)=\textbf{k}(t)/\text{k}$. The fibre bundle linking a phase to a parameter transformation was introduced by Simon
\cite{Simon1983}. The adiabatic geometric phase can then be calculated from the connection on Simon's fibre bundle. The adiabatic geometric phase $\phi_{g,a}$ acquired by a photon when the direction of the wave vector is cycled reads: 
\begin{equation}\label{eq8}
\phi_{g,a}=-\sigma\Omega_{k}(\mathcal{C}), 
\end{equation}
where $\Omega_{k}(\mathcal{C})$ is the solid angle subtended by the path formed on the sphere of directions $\textbf{k}$ and where the helicity $\sigma$ denotes the projection of the spin onto $\textbf{k}$, which takes values $\sigma=1$ for left-handed and $\sigma=-1$ for right-handed circularly polarized light. Here, $\phi_{g,a}$ is analogous to a well-known adiabatic phase, namely the Berry phase obtained from the evolution of a spin particle interacting with a time-varying magnetic field $\textbf{B}(t)$ of constant amplitude, where the directions of the magnetic field, $\textbf{R}(t)=\textbf{B}(t)/B$ are the parameters \cite{Berry1984}. 

The geometric phase $\phi_{g,a}$ produces a characteristic effect: when $\textbf{k}$ recovers its initial orientation, the polarization axis of linearly and elliptically polarized light is rotated. This rotation can be understood in terms of circular birefringence:  The left and right circularly polarized light components of the beam
acquire opposite geometric phases \cite{TomitaChiao1986}.

\begin{figure}[tb]
  \centering
  \includegraphics[width=\linewidth]{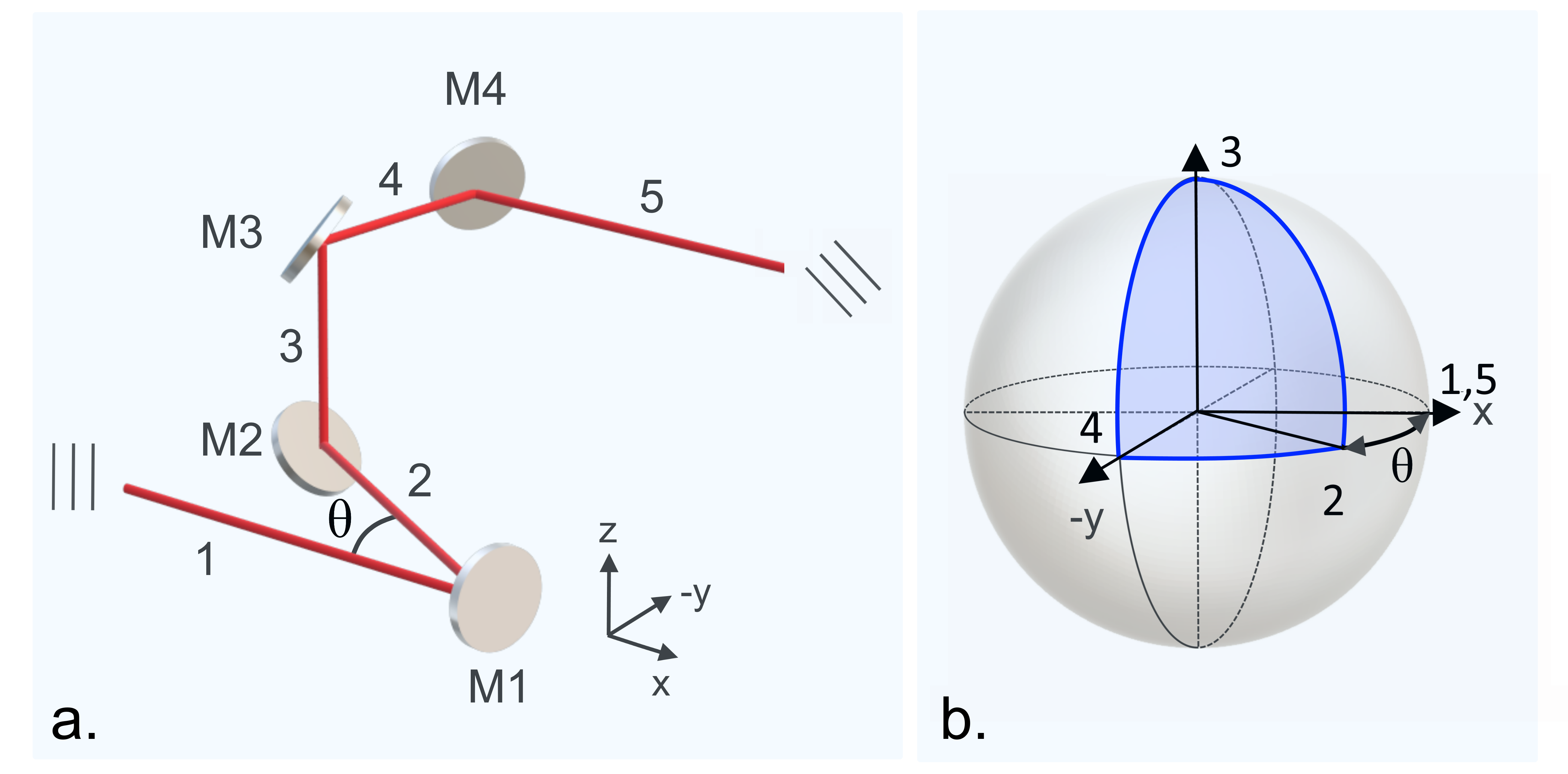} 
\caption{Generation of SR phases. a. A linearly polarized beam is taken along a non planar trajectory, the mirrors M2 and M3 form a beam elevator. At the end of the trajectory the polarization axis is rotated. b. Path formed on the sphere of spin directions.
}
\label{F5}
\end{figure}

It soon appeared that this rotation can also be observed when light is redirected by a sequence of mirrors \cite{Kitano1987,BerryNature1987}. However, in this case the transformation is non-adiabatic because mirror reflections reverse the helicity. An attempt was made to continue using the parameter space to calculate the geometric phases but it became clear that this description had reached its limits, it had to rely on modified wave-vectors and had to account for occasional $\pi$ phase shifts \cite{Kitano1987}. At a similar time, Aharonov and Anandan changed their emphasis from the parameter space to the state space. They showed that adiabaticity is not a necessary condition for the existence of geometric phase, but it is the state evolution that matters \cite{AA1987}. Similarly, for the case of a spin particle in a magnetic field Anandan considered 
the evolution of the spin instead of the evolution of the magnetic field direction (parameter), hereby lifting the adiabatic requirement \cite{Anandan1992}. For spin $1/2$ particles, the sphere of spin directions is the Bloch sphere. Inspired by this work, Chiao et al. shifted the emphasis from the evolution of the direction of the wave vector to the evolution of the spin vector $\textbf{S}$ of photons \cite{Chiao1988}.  The geometric phase, now called spin-redirection (SR) phase, became:
\begin{equation}\label{eq9}
\phi_{g,a}=-\sigma\Omega_{\rm SR}(C),   
\end{equation}
where $\Omega_{\rm SR}(\mathrm{C}) $ is the familiar solid angle formed on the sphere representing the directions of the spin $\textbf{S}$ of the photon in real space. Fig.~\ref{F5} illustrates how a beam of light can be taken along a non-planar trajectory using a succession of mirrors, and shows the respective path traced on the sphere of spin directions of photons. 

Note, however, that unlike the Bloch sphere of spin-$1/2$ particles, which incorporates information on the direction of the spin in real space and identifies all pure states, the sphere of spin directions of photons is not a state space. The state space identifying all pure polarization states was presented earlier - it is the Poincar\'e sphere. However, the Poincar\'e sphere is built on the assumption that polarization characterizes the oscillation of a two dimensional electric field contained in the plane transverse to a constant propagation direction. If the propagation direction varies, so does the orientation of the transverse plane spanning the polarization.

When the propagation direction of a beam of light is varied, the electric field becomes a 3-component vector $\textbf{E}=(E_{x},E_{y},E_{z})$ in the laboratory frame. The normalized state vector $\ket{\psi}$ representing the system then corresponds to a rotated 3-component spinor, reflecting the spin-1 nature of photons \cite{Berry1987,Hannay_1998}. 
We are thus dealing with a 3-state system, of state space $\mathbb{C}\text{P}^{2}$, which is no longer an ordinary sphere. State spaces become difficult to visualize as their dimension increases, and so does picturing the evolution of the state in that space. Fortunately, Majorana provided an elegant way to circumvent this difficulty.
Majorana was studying the behaviour of a spin system of arbitrary angular momentum $\mathbf{j}$ in the presence of a magnetic field when he realized that varying the direction and magnitude of the magnetic field amounts to rotating  $\mathbf{j}$ \cite{Majorana1932}. After the rotation, a system that was originally in an eigenstate finds itself in a superposition of $2j+1$ states. The $\mathbf{j}$-spin problem thus becomes equivalent to relating angular momentum states associated with different directions in space \cite{Schwinger1977}. This is equivalent to the problem we encounter when we wish to
compare polarization along a varying propagation direction. 
Majorana went on by representing a spin $\mathbf{j}$ state as a constellation of $2j$ points on an ordinary sphere. Each point, poetically called a star, represents the direction of a spin-$1/2$ angular momentum \cite{BlochRabi1945}. From a geometric perspective, what Majorana really did was write an $n$-dimensional state space $\mathbb{C}\text{P}^{n}$, as an unordered product of $n-\mathbb{C}\text{P}^{1}$, the space of all unordered sets of $n$-points on a sphere. 

\begin{figure}[tb]
  \centering
  \includegraphics[width=0.8\linewidth]{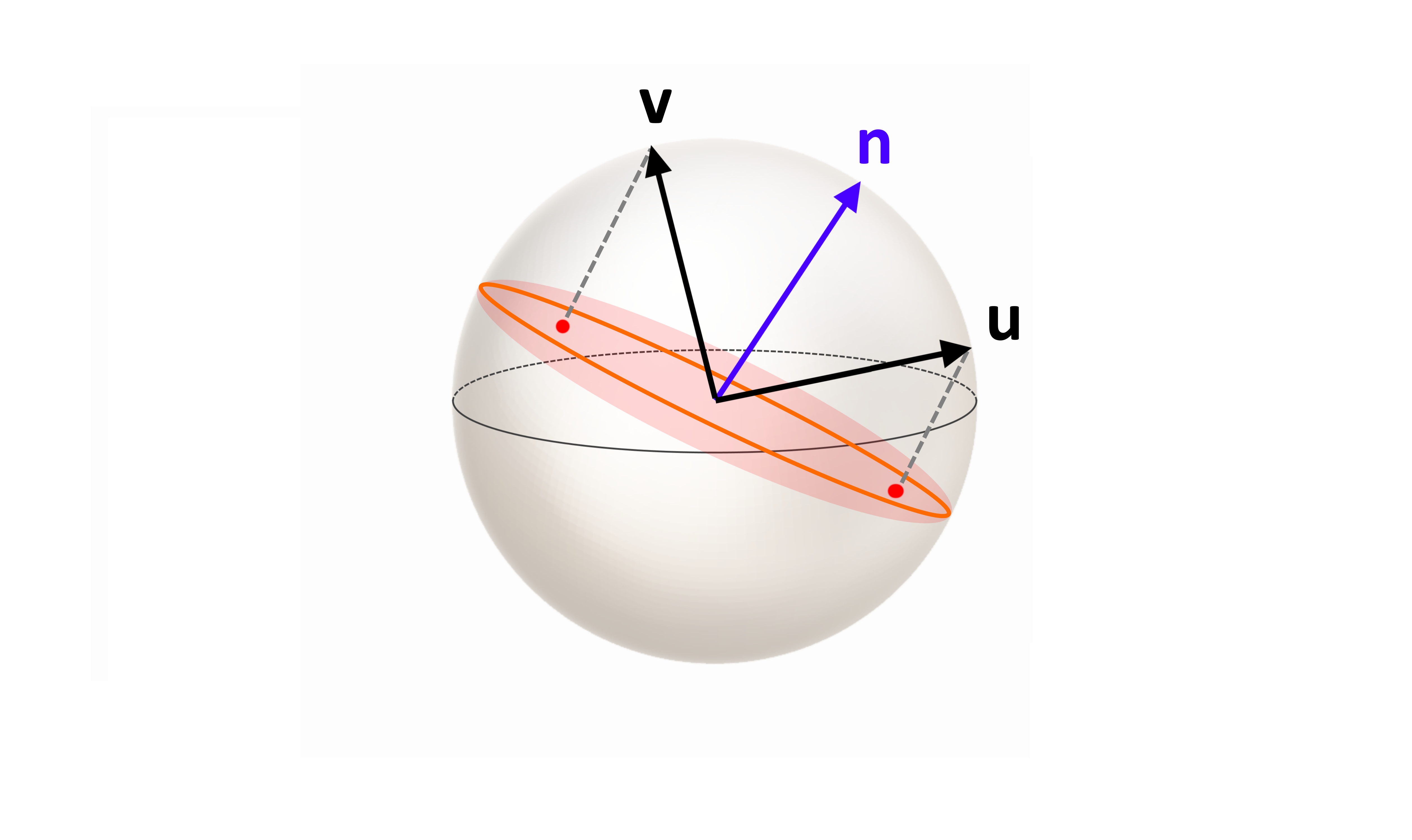} 
\caption{Hannay's representation of 3D polarized light. The stars correspond to the tips of the vectors $\textbf{v}$ and $\textbf{u}$. The polarization ellipse is represented in orange (solid gray circle), $\text{n}$ is aligned with the propagation direction. The foci of the ellipse correspond to the projection of the stars onto the plane orthogonal to their bisector.}
\label{f6}
\end{figure}

In 1998, Hannay used the Majorana representation to visualize 3D polarized light as two stars on a sphere (see Fig.\ref{f6}) \cite{Hannay_1998}. In his work, he managed to relate the Majorana's mathematical construct with a concept of a polarization ellipse and its orientation direction in 3D space, familiar to all researchers working in optics. Specifically, he  showed that the foci of the polarization ellipse are given by the projection of the stars onto the plane perpendicular to the bisector of their angle.  
He also deduced the geometric phase associated with the transformation of 3D polarized light from the circuits traced by the two stars. 

Nonparaxial fields, for which the electric field component along the propagation direction is non-negligible, have attracted increasing attention in the past decades, in virtue of their capacity to mix the spin and orbital angular momentum content of the beam \cite{BARNETT1994,BliokhAlonso2010,Ma2016}. This renewed the interest in their geometric phases and has brought the Majorana representation back in the spotlight \cite{Bliokh2019,Alonso2020}.

While picturing the state evolution is certainly helpful, we have shown in section IV that all we really need to calculate the geometric phase is the connection on the relevant fibre bundle. For a spin-1 system, the base space is the state space $\mathbb{C}\text{P}^{2}$, the total space is the Hilbert space $\mathrm{H}_{3}$ and the fibre is $\mathrm{U}(1)$. A good description of this fibre bundle is provided in \cite{Bouchiat1988}. In this case, the set of pure states is characterized by four parameters consisting of three Euler angles $\theta,\varphi,\alpha$ giving the orientation in space of the principal axis of the polarisation vector, and an extra parameter defining the shape of the ellipse $\delta$. The geometric phase of 3D polarized light reads:
\begin{equation}\label{Bouchiat}
\phi_{g}=\oint_{C}\mathcal{A}=\oint_{C}(\mathrm{sin}\delta\,\cos\theta\,\mathrm{d}\varphi+
(\mathrm{sin}\delta-1)\,\mathrm{d}\alpha). 
\end{equation}
Hannay recognized that this expression is equivalent to the one found using the Majorana representation \cite{Hannay_1998}. The set of coordinates on $\mathbb{C}\text{P}^{2}$ contains some singularities, like the ones we identified at the poles on the Poincar\'e sphere \cite{Bouchiat1988}. It would be interesting to study whether this has physical consequences. A clear geometric interpretation of the limiting cases, where the geometric phase becomes the Pancharatnam-Berry phase or the spin-redirection phases would also be useful. It has been suggested that, in the context of general relativity, $\mathbb{C}\text{P}^{2}$ can be regarded as a half pseudoparticle surrounded by a cosmological event horizon and that it shares properties of the Yang-Mills instanton \cite{Gibbons1978}. We ask whether investigating the phase holonomies of 3D polarized light could be exploited to study these systems. 

Turning paraxial light into a 3D field is relatively straightforward, one can use a high numerical aperture or rely on scattering \cite{BliokhElena2011}, measuring the entire electric field however, is a highly challenging task. 
Fortunately, it is now possible to access high dimensional state space without breaking the paraxiality, by structuring light in its spatial degree of freedom.  

\subsection{Geometric phases of spatial transverse modes}

Optical modes are characterized not only by their polarization but also by their spatial profile, determining both phase and intensity distribution across the beam \cite{Forbes2021}. While polarization is usually limited to a two dimensional state space, there is an infinite number of orthogonal spatial modes, with Hermite-Gaussian ($\mathrm{HG}_{n,m}$) and Laguerre-Gaussian ($\mathrm{LG}_{p}^{\ell}$) modes providing possible basis sets in Cartesian and polar coordinates respectively. 

A spatial transverse mode of order $\text{N}=n+m=2\;p +|\ell|$, may be represented by a normalized vector $\ket{\psi}$ which may refer to a coherent state as approximated by a classical light beam, or the wavefunction of a photon. The state vector then lives in a Hilbert space of dimension $\text{N}+1$, and the state space is $\mathbb{C}\mathrm{P}^{\text{N}}$ \cite{Courtial1999}.

\begin{figure*}[t]
  \centering
  \includegraphics[width=\textwidth]{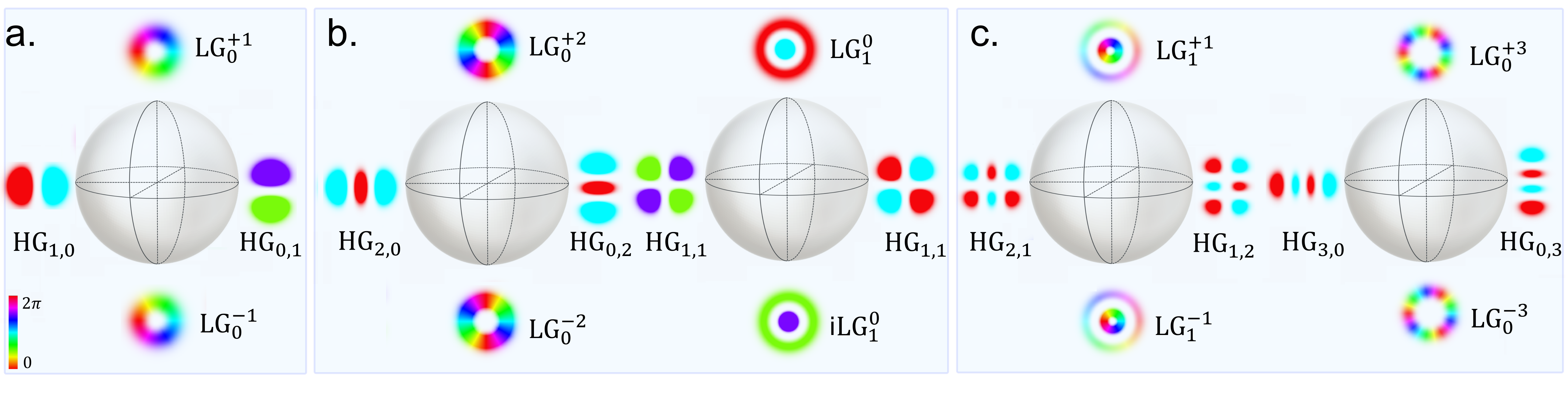}
\caption{Spheres of spatial transverse modes following the convention of \cite{Habraken2010}, with all modes on one sphere linked by optical mode converters and rotators. a. The sphere of first order modes is the direct analogue of the Poincar\'e sphere. b. and c. Second and third order modes each are represented on two spheres.}
\label{f7}
\end{figure*}

For $\text{N}=1$, $\ket{\psi}$ is a two-state system, of the form of Eq. \ref{eq0}, where $\ket{0}$ and $\ket{1}$ correspond to any two linear dependent orthogonal modes, conveniently we can choose ${\rm LG}_0^1$ and ${\rm LG}_0^{-1}$ modes or ${\rm HG}_{0,1}$ and ${\rm HG}_{1,0}$ modes. Like for all two-state systems, this state space can be pictured as an ordinary sphere, the so-called \textit{sphere of first order modes}, shown in Fig.~\ref{f7}.a \cite{VANENK199359,Padgett:99,Agarwal1999}. By convention, the poles represent the modes ${\rm LG}_0^{\pm 1}$ and the equator corresponds to first order HG modes of varying alignment. All diametrically opposed modes form a suitable orthogonal basis system, from which all modes on the sphere can be obtained as a linear superposition. A path $\mathrm{C}$ can be formed on the sphere using a sequence of mode-preserving optical elements, like a pair of Dove prisms acting as a mode rotator, or a pair of cylindrical lenses acting as a mode convertor \cite{BEIJERSBERGEN1993123}. The geometric phase associated with the transformation of first order modes reads as \cite{VANENK199359,Galvez2003}:
\begin{equation}
\phi_{g,\text{N}=\text{1}}=-\frac{1}{2}\Omega(C),
\end{equation}
where $\Omega(C)$ is the solid angle enclosed by the path formed on the sphere, in analogy to the PB phase.  This is not surprising since the underlying geometry is the same. The phase $\phi_{g,\text{N}=\text{1}}$ can be interpreted as the holonomy of the connection on the Hopf fibration, where the base space corresponds to the sphere of first order modes. This interpretation, to the best of our knowledge, has not yet been made explicit in the literature.

For $\text{N}>1$, the dimension of the state space $\ket{\psi}$ grows.  
Second order modes, for example, need to be expressed in terms of not two, but three fundamental modes, where $\text{LG}_{0}^{+2}$, $\text{LG}_{1}^{0}$ and $\text{LG}_{0}^{-2}$ form a complete basis; third order modes require four fundamental modes, where $\text{LG}_{0}^{+3}$, $\text{LG}_{1}^{+1}$, $\text{LG}_{1}^{-1}$ and $\text{LG}_{0}^{-3}$ form a complete basis, and so on. So far, geometric phases have been calculated on two-dimensional subspaces of these high dimensional state spaces, represented as spheres.
In \cite{Habraken2010}, a number of $(\text{N}+1)/2$ spheres is used to represent modes of odd mode order $\text{N}$ and a number of $(\text{N}+2)/2$ spheres in order to represent modes of even mode order. In practice, this means that both second order and third order modes will be represented using two ordinary spheres (see Fig.\ref{f7}. b and c.).

However, unlike for first and third order modes, not all the poles of the spheres of second order modes carry orbital angular momentum, indeed, one sphere presents the $\text{LG}_{1}^{0}$ mode and the mode ($i \text{LG}_{1}^{0}$) at the poles.
This is a general feature of even modes. Also note that the modes at the equator of the spheres no longer correspond to the linear superposition of the poles, as would be the case for generalized Poincar\'e spheres: for the first sphere of second order modes, for example, we would expect to see $\mathrm{HG}_{1,1}$ rather than $\mathrm{HG}_{0,2}$ modes at the equator. This reflects the choice of the authors in \cite{Habraken2010} to obtain all the modes on these spheres by performing a mode-preserving transformation on the modes at the poles, which can easily be realized in the laboratory using astigmatic mode converters (to move along lines of constant longitude) and image rotators (to move along lines of constant latitude).
The geometric phase obtained from a cyclic mode-preserving transformation, which effectively forms a path on these sub-dimensional state spaces, is then calculated using \cite{Calvo2005}:
\begin{equation}\label{calvoeq}
\phi_{g,\text{N}}=-\frac{1}{2}\ell\,\Omega,  
\end{equation} 
where $\Omega$ is the solid angle formed on the sphere describing the transformation. Interestingly, when a path is formed on a sphere on which all modes carry the same amount of orbital angular momentum, like the second sphere of second order modes, no geometric phase is generated \cite{Galvez2005}. This would indicate that geometric phases are mediated by a variation of orbital angular momentum, in the same way as polarization transformations that generate a PB phase involve variation of the spin angular momentum \cite{VANENK199359,Tiwari1992,2109.10169}.
While the sphere-based representation is useful as it directly relates to transformations that are easily realisable in the laboratory, it is not suitable to describe generic transformations in the state space of higher order modes.

Interpreting spatial transverse modes in terms of a fibre bundle would allow us to explore geometric phases over the entire state space, not just over two-dimensional sub-spaces. For a spatial transverse mode of order $\text{N}$, the relevant fibre bundle would be the so called tautological line bundle, with the same base space $\mathbb{C}\mathrm{P}^{\mathrm{N}}$, total space $\mathrm{H}_{\mathrm{N+1}}$ and $\mathrm{U}(1)$ as the fibre. It would be interesting to determine, at least theoretically, whether transformations over extended portions of the state space lead to the discovery of new geometric or topological phases, with possible applications for topological photonics and quantum communication. In tandem with such fundamental discussions we may develop experimental techniques that can realize general forms of mode transformations, hereby leading to the expansion of the  spatial mode shaping toolbox.  

In reality, the exploration of high order spaces in optics has already begun. Indeed, a Majorana representation of structured Gaussian beams been introduced in 2020, revealing that geometric phases born from cyclic model transformation of generalised structured-Gaussian beams can be discrete \cite{Cuevas2020}. Following Hannay's observation about the Majorana represention
we expect this result to be confirmed by fibre bundle theory. Investigations are still at an early stage and other geometric and topological phases may still be waiting to be discovered. Also, note that more general mode solutions of the paraxial wave equation have received increasing attention in the past years and promise to expand the horizon of geometric phases even further \cite{Alonso:17,Dennis2017,Dennisalonso_2019}. 

Knotted beams, for which the locus of phase singularities form linked and knotted threats upon propagation \cite{Leach2004,Berry2001} may also uncover intriguing geometric phases.

So far, we have considered the spatial and polarization degree of freedom of light independently. We shall now study vector light fields where they become nonseparable, and discuss the implications with regard to their geometric phases.

\subsection{Geometric phases of general vectorial fields}

\begin{figure}[tb]
  \centering
  \includegraphics[width=1.0\linewidth]{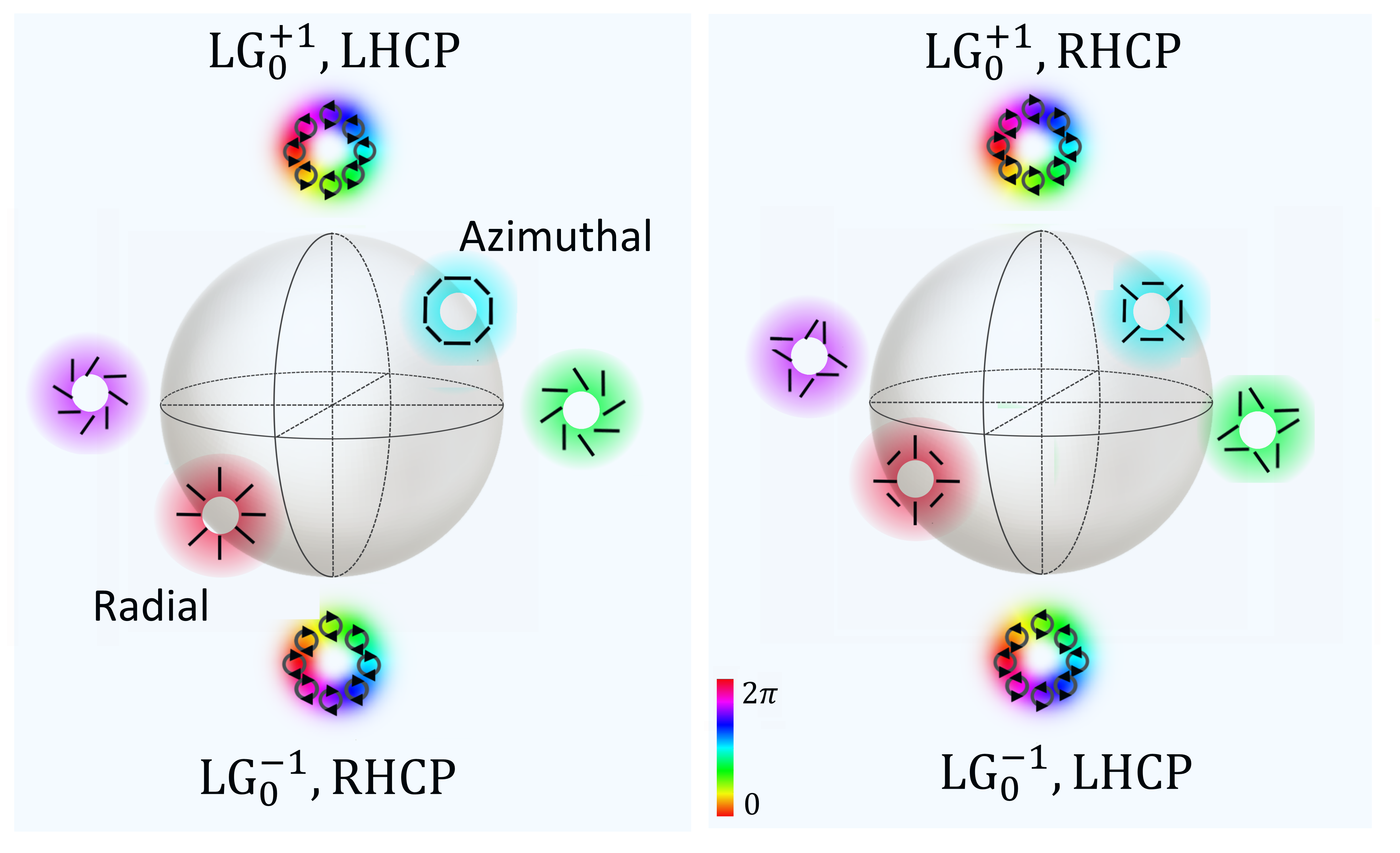} 
\caption{Spheres of first order polarized modes. On the left sphere, $\sigma=-\ell$ at the poles and the modes on the equator are co-rotating vector modes. For the sphere on the right, $\sigma=\ell$ and the modes on the equator are counter-rotating modes.}
\label{F8}
\end{figure}

Combining the polarization and spatial degrees of freedom of light amounts to building a bipartite system, where the Hilbert space of the system corresponds to the tensor product of the individual spaces $\mathrm{H_{pol}} \otimes \mathrm{H_{spa}}$.
For simplicity, we shall only consider first order transverse modes, in which case we are dealing with a two-qubit system \cite{Khouri2007}. Homogeneously polarized light is described by a product state, separable into a qubit that describes the polarization, and one for the spatial mode. Light with non-homogeneous polarization instead is non-separable in these distinct degrees of freedom \cite{Khouri2007}. Well-known examples of non-separable modes are radially and azimuthally polarized modes of the form $ \mathrm{LG}_{p}^{1} \sigma_{\pm}+\mathrm{LG}_{p}^{-1} \sigma_{\mp}$, where $\sigma_\pm$ represents left and right circular polarised light respectively \cite{Zhan:09,Otte2016,Liu2018,Selyem2019}. These modes have received increasing attention as they can be focused to tighter spots than their uniformly polarized counterparts \cite{Youngworth:00}.
General vector beams built from first order modes are usually represented using two Poincar\'e-like $\mathrm{S}^{2}$ spheres, shown in Fig.~\ref{F8}, where the poles correspond to uniformly circularly polarized vortex modes, of helicity $\sigma=\pm 1$ and carrying an optical vortex of topological charge $\ell=\pm 1$. The states on the equator correspond to co-rotating modes, such as radial and azimuthal modes, and counter-rotating modes \cite{Holleczek:11,Milione2011}. The geometric phase associated with the transformation of these modes is then calculated from the solid angle $\Omega$ formed on the relevant sphere \cite{Milione2011}:
\begin{equation}
\phi_{g}=\pm\frac{1}{2}(\ell+\sigma)\Omega,  
\end{equation}
The total geometric phase is thus linked to the total angular momentum of the beam $\ell+\sigma$. This was experimentally verified in \cite{Milione2012}, where a combination of a half wave plate and an astigmatic mode converter realized the mode transformation.

Again, the space of pure states of a two-qubit system, $\mathbb{C}\mathrm{P}^{3}$, is not an ordinary sphere. However, depending on the degree of separability of the states, the associated substate space may take a more recognizable form \cite{bengtsson_zyczkowski_2006}. 
Separable states for instance, form a $\mathbb{C}\mathrm{P}^{1} \otimes \mathbb{C}\mathrm{P}^{1}$ subspace, called Segre embedding \cite{bengtsson_zyczkowski_2006}.
There is a curious correspondence between the geometry of arbitrary separable states and the fibre bundle with base space $\mathrm{S}^{4}$, fibre $\mathrm{S}^{3}$ and total space $\mathrm{S}^{7}$ \cite{Mosseri_2001}. Interestingly, this fibration is a generalization of the Hopf fibration, and is normally used to describe quaternions, but it has also been used to study the geometric phases of two-qubit systems \cite{L_vay_2004}. The phase associated with the cyclic evolution of a maximally entangled state is purely topological \cite{MosseriMilam,L_vay_2004}. The topological phase arising under the cyclic transformation of maximally non-separable optical modes has been measured in \cite{Khouri2007,Souza2014,Matoso2019}. It would be useful to study whether the tautological line bundle over $\mathbb{C}\mathrm{P}^{3}$ yields similar results, considering that the Hopf fibration is not originally intended for the description of complex fields, and does not generalize to arbitrary dimensions.   

In this section, we have considered vector modes built from first order modes, but a more general description would include arbitrary vector fields based on spatial modes of higher order. The sphere-based representation presented in Fig.~\ref{F8} then needs to be expanded by allowing the LG beams at the poles to be of different topological charge, $\ell$ and $m$. The associated geometric phase then reads \cite{Yi2015}:
\begin{equation}
\phi_{g}=-\frac{\ell-(m+2\sigma)}{4}\Omega,
\end{equation}
where $\Omega$ is the solid angle formed on the modified sphere under consideration. This phase was measured in \cite{Liu:17} using two identical q-plates. Interpreting this phase in terms of a fibre bundle is certainly possible, but would be pure speculation without first addressing the questions raised by two-qubit systems.

\section{Summary and perspectives}

Fibre bundle theory presents a rigorous treatment for the understanding of phases. It sheds light on the origin of the solid angle law linking a geometric phase to the path formed on a generalised Poincar\'e sphere representing the mode space when the mode is transformed. These spaces however often only represent two-dimensional subspaces of a high dimensional state space. They cannot be represented by a sphere and are difficult to visualize, they may however present geometric and topologic features giving birth to interesting geometric and topologic phases, undetectable in the two-dimensional sub-space descriptions. Majorana-based representations are slowly emerging, they are capable of providing an accurate expression for geometric phases born in high dimensional state spaces, while providing a clear visual interpretation. At a more fundamental level, tautological line bundles should be used to calculate these geometric phases, the only ingredient needed is the connection on these bundles. Research on general vectorial modes raises the question of how non-separability can be accounted using fibre bundles, and whether this causes measurable effects.      

With this colloquium we hope to encourage collaborations between the optics and the mathematics communities, as we believe that higher order structured Gaussian modes and vector modes may allow the exploration of new concepts.       

\section{Acknowledgements}

The authors acknowledge insightful discussions with Remy Mosseri, Antonio Zelaquett Khoury, Kerr Maxwell and Mark Dennis.

This work was supported by the Royal Society through a Newton International Fellowship (NIF/R1/192384), the Leverhulme Trust and the UK's Engineering and Physical Research Council with grant number EP/V048449/1.

\bibliography{sample}
\end{document}